# ИНФОРМАТИКА

УДК 004.9:66.013.512

## МОДУЛЬНАЯ ТЕХНОЛОГИЯ РАЗРАБОТКИ РАСШИРЕНИЙ САПР: МОЛНИЕЗАЩИТА ЗДАНИЙ И СООРУЖЕНИЙ

В.В.Мигунов, Р.Р.Кафиятуллов, И.Т.Сафин

*ЦЭСИ РТ при КМ РТ, г. Казань*

Аннотация

Модульная технология разработки проблемно-ориентированных расширений САПР применена к задаче проектирования молниезащиты зданий и сооружений с реализацией в программной системе TechnoCAD GlassX. Разработана системная модель проектов молниезащиты, включающая структурированное параметрическое представление (свойства объектов и их связи, общие установки и установки по умолчанию) и операции над ним, автоматизирующие проектирование.

Библ.2

Abstract

V.V. Migunov, R.R. Kafiyatullov, I.T. Safin. The modular technology of development of the CAD expansions: protection of the buildings from the lightning // Izvestiya of the Tula State University/ Ser. Mathematics. Mechanics. Informatics. Tula: TSU, 2003. V._. N _. P. __–__.

The modular technology of development of the problem-oriented CAD expansions is applied to a task of designing of protection of the buildings from the lightning with realization in program system TechnoCAD GlassX. The system model of the drawings of lightning protection is developed including the structured parametric representation (properties of objects and their interdependence, general settings and default settings) and operations with it, which efficiently automate designing.

Bibl.2

Настоящая работа посвящена применению модульной технологии разработки проблемно-ориентированных расширений систем автоматизированного проектирования (САПР) реконструкции предприятия, общие положения которой изложены в [1]. Объект приложения технологии - автоматизация подготовки чертежей проектов молниезащиты зданий и сооружений (МЗС), выполненная в САПР TechnoCAD GlassX. Кроме обычных требований единой системы конструкторской документации и системы проектной документации для строительства, чертежи МЗС должны отвечать правилам [2]. В [2] задаются методы расчета горизонтальных и вертикальных сечений зон защиты молниеприемников различных типов. Изображение и характеристика этих сечений на чертеже защищаемых зданий и сооружений и является проектом их молниезащиты.

Учитываются и рассчитываются комплексы из стержневых (одиночных, двойных и многократных), тросовых (одиночных и двойных) и сеточных молниеприемников (МП). Основой автоматизации является быстрый (мгновенный) расчет зон защиты по [2] для подбираемых проектировщиком вариантов молниезащитных сооружений с демонстрацией на чертеже нескольких сечений зон защиты в нескольких проекциях и таблицы основных данных ("Таблица расчета молниезащиты").

Сечения зон защиты всегда простые, без изломов секущих плоскостей. Все вертикальные сечения – вынесенные (располагаются вне изображения предмета). На виде сверху располагаются одно или несколько горизонтальных сечений зон защиты для разных высот, на каждом вертикальном сечении – только одно сечение зон защиты. Каждое из сечений зон защиты может учитывать свой набор молниеприемников.

При разработке МЗС используется заранее подготовленный чертеж с изображениями зданий и сооружений (например, строительная подоснова зданий), для которых разрабатывается МЗС. Чертеж МЗС представляется в виде одного обязательного вида сверху (плана), нескольких необязательных вертикальных сечений и содержит следующие специфичные элементы:

- изображения зданий и сооружений в нескольких видах, опор молниеотводов и токоотводов, подземные инженерные сооружения: подземный трубопровод, резервуары, фундамент и пр. Эти элементы вычерчиваются обычными средствами TechnoCAD GlassX, без автоматизации в рамках данного проблемно-ориентированного расширения;
- изображения стержневых молниеприемников. Отдельно стоящий на земле МП на виде сверху изображается квадратом с диагональными линиями; стоящий на каком-либо объекте МП на виде сверху изображается жирной точкой; на других проекциях и видах МП изображаются равнобедренным треугольником, вытянутым вверх по его высоте;
- изображения сеточных МП в виде ограничивающих многоугольников. При разработке МЗС

многоугольники для наглядности штрихуются как неметаллы (2x45°), при дооформлении чертежа они штрихуются проектировщиком более точно, с учетом шага сетки на различных участках;
- изображения одиночных и двойных тросовых МП;
- позиционные обозначения МП;
- обозначение вида сверху;
- обозначения вертикальных сечений на виде сверху (включая изображение секущего отрезка) и на самом вертикальном сечении;
- размеры, характеризующие расстояния между МП;
- линии горизонтальных и вертикальных сечений зон защиты, порождаемых различными наборами МП;
- текстовые обозначения горизонтальных сечений зон защиты на виде сверху с указанием высот;
- размеры радиусов зон защиты одиночных МП на нулевой высоте;
- размеры минимальной ширины зон защиты двойных МП на нулевой высоте;
- таблица расчета молниезащиты;
- изображения заземлителей, обеспечивающих растекание тока молнии в земле;
- другая текстовая информация, более общего характера, относящаяся к зонам молниезащиты, конструкции МП, обозначениям имеющихся на чертеже видов.

Используется внутренняя трехмерная прямоугольная декартова правая система координат с единицами измерения мм натуры, начало которой считается находящимся на нулевой отметке высоты и называется далее базовой точкой. На виде сверху оси X и Y внутренней системы координат идут вдоль одноименных осей чертежа, а ось Z идет от плоскости чертежа к пользователю.

Параметрическое представление содержит списки объектов МЗС, установки построения по умолчанию и общие установки МЗС, вводимые, выбираемые или указываемые проектировщиком. Для краткости ниже словами "установки шрифта" будут называться размер, угол наклона и степень сжатия чертежного шрифта системы TechnoCAD GlassX. Все координаты и размеры задаются во внутренней системе координат, если не оговаривается иное. Если речь идет о размерах в отпечатанном чертеже, говорится о системе отсчета Бумага.

*Молниеприемники*

Список молниеприемников для каждого МП содержит: позиционное обозначение (однострочный текст); тип (однострочный текст); вариант конструкции (стержневой, сеточный, тросовый и двойной тросовый); признак того, что стержневой МП – отдельно стоящий, а не установлен на каком-либо объекте; высота конструкции (кроме сеточного); список трехмерных точек, цвет и тип линии. Для стержневого варианта задается одна точка – верхняя точка стержня, для сеточного – не менее трех вершин ломаной без самопересечений с обходом против часовой стрелки, ограничивающей плоский горизонтальный многоугольник сетки (координаты Z одинаковы для всех углов многоугольника), для тросового – две точки опор на их высоте (для обеих точек координаты Z одинаковы), для двойных тросовых – то же плюс смещение второго троса от первого и его высоту. Позиционное обозначение и высота МП размещаются в таблице расчета молниезащиты соответственно в графах "N°N° молниеприемников" и "Высота молниеприемника (h)";

*Вертикальные сечения чертежа*

Список (вертикальных) сечений чертежа – проекций для каждого сечения содержит: буквенное обозначение сечения (возможно, включая слово "Повернуто"); масштаб Бумага/Натура; проекцию базовой точки внутренней системы координат на плоскость сечения в системе отсчета Бумага всего чертежа; секущий отрезок на плоскости XY внутренней системы координат (след секущей плоскости); параметры для вычерчивания буквенного обозначения вертикального сечения на виде сверху: длина штрихов на краях; смещение стрелки от края штриха; смещение центра габаритов буквенного обозначения от ближайшего конца секущего отрезка (все размеры в мм Бумаги); параметры для вывода буквенного обозначения сечения чертежа на своем виде: смещение точки середины полки буквенного обозначения относительно проекции базовой точки в системе отсчета Бумага (текст обозначения центрируется относительно нее); смещения от полки до текстов над и под ней; вариант включения масштаба сечения в его буквенное обозначение (вывести в одной строке; вывести отдельной строкой под полкой); установку шрифта. Положение буквенного обозначения относительно секущего отрезка задает направление взгляда. На каждом вертикальном сечении чертежа располагается одно сечение зоны защиты.

*Сечения зон защиты*

Список сечений зон защиты (горизонтальных и вертикальных) для каждого сечения содержит:

ссылку на вертикальное сечение чертежа из их списка (в том числе 0 для вида сверху); список молниеприемников, учитываемых при построении сечения (не указанные в этом списке МП не изображаются на сечении чертежа и их влияние на сечение зоны защиты игнорируется); отметку высоты сечения (для горизонтальных сечений зон защиты, размещаемых на виде сверху); цвет и тип линии границы сечения зоны защиты.

*Тексты молниеприемников*

Список текстов молниеприемников для каждого текста содержит: ссылку на МП из их списка; ссылку на сечение чертежа из их списка (для текстов, располагающихся на вертикальном сечении); смещение точки начала текста относительно проекции базовой точки в системе отсчета Бумага; смещение точки указания сноски от первой точки МП в системе отсчета Бумага; признак направления сноски на конец или на начало полки.

*Тексты сечений зон защиты на виде сверху*

Список текстов (отметок высот) сечений зон защиты на виде сверху для каждого текста содержит: смещение точки начала текста от проекции базовой точки в системе отсчета Бумага; ссылку на элемент списка сечений зон защиты; угол направления сноски из точки начала текста; признак направления сноски на конец или на начало полки; признак расположения текста в двух строках, а не в одной.

*Размеры расстояний между осями стержневых молниеприемников*

Список (линейных) размеров – расстояний между осями стержневых молниеприемников в метрах для каждого из них содержит: ссылки на два МП из их списка, расстояние между которыми указывается в размере; ссылку на вертикальное сечение чертежа из их списка (для размеров, расположенных на вертикальных сечениях); смещение размерной от первой точки начала выносных и смещение текстов от размерной линии в направлении допустимого расположения текстов (система отсчета Бумага); сведения о засечках или стрелках (стрелка внутрь, стрелка наружу, засечка).

*Размеры радиусов зон защиты одиночных стержневых молниеприемников на виде сверху*

Список размеров радиусов (в метрах) зон защиты одиночных стержневых молниеприемников на виде сверху (представляют собой линейные размеры типа "Радиус" с текстом типа "Rx1 = 17.00") для каждого такого размера содержит: текст обозначения параметра, например "Rx1"; ссылка на МП из их списка (начало размерной); ссылка на сечение зоны защиты из их списка; угол направления размерной против часовой от положительного направления оси X чертежа; признак автоматического позиционирования текста размера посередине размерной линии; смещение текстов от размерной линии в направлении допустимого расположения текстов в системе отсчета Бумага (учитывается при автоматическом позиционировании текста размера); положение точки начала текста относительно базовой точки в системе отсчета Бумага (учитывается при ручном позиционировании); признак наличия сноски (отсутствует (по умолчанию), от середины размерной, от начала или конца размерной); признак проведения сноски к началу (по умолчанию) или концу полки; сведения о засечках или стрелках (стрелка внутрь, стрелка наружу или засечка).

*Размеры радиусов зон защиты одиночных молниеприемников на вертикальных сечениях*

Список размеров (в метрах) радиусов зон защиты одиночных молниеприемников на вертикальных сечениях на нулевой отметке высоты (представляют собой линейные размеры с текстом типа "R1 = 25.00") для каждого такого размера содержит: текст обозначения размера – радиуса, например, "R1"; ссылку на молниеприемник из их списка (начало размерной); ссылку на вертикальное сечение из их списка; смещение размерной от первой точки начала выносных и смещение текстов от размерной линии в направлении допустимого расположения текстов (оба параметра в системе отсчета Бумага); сведения о засечках или стрелках (стрелка внутрь, стрелка наружу или засечка); направления размерной от МП (влево или вправо).

*Размеры минимальных ширин зон защиты двойных молниеприемников на виде сверху*

Список размеров (в метрах) минимальных ширин зон защиты двойных молниеприемников на виде сверху (представляют собой линейные размеры типа "Радиус" с текстом типа "Rcx = 13.50") для каждого размера содержит: текст обозначения параметра, например "Rcx"; ссылки на два МП из их списка (их порядок важен: размерная идет налево перпендикулярно направлению от первого МП ко второму); ссылка на сечение зоны защиты из их списка; признак автоматического позиционирования текста размера посередине размерной линии, а не вручную пользователем; смещение текстов от размерной линии в направлении допустимого расположения текстов в системе отсчета Бумага (учитывается при автоматическом позиционировании текста размера); положение точки начала текста относительно базовой точки в системе отсчета Бумага (учитывается при ручном позиционировании); признак наличия сноски (отсутствует (по умолчанию), от середины размерной, от начала или конца размерной); признак проведения сноски к началу (по умолчанию) или концу полки; сведения о засечках или стрелках (стрелка внутрь, стрелка наружу или засечка).

*Молниеприемники, вносимые в таблицу расчета молниезащиты*

Список молниеприемников, вносимых в таблицу расчета молниезащиты, для каждого элемента списка содержит информацию для одного одиночного или двух связанных друг с другом (двойной) МП: ссылку на молниеприемник из их списка (это либо одиночный молниеприемник, либо первый из пары молниеприемников, составляющих двойной МП); ссылка на второй молниеприемник (для двойного МП); значение отметки высоты защищаемого уровня относительно нулевой отметки.

*Заземлители на виде сверху*

Список заземлителей для каждого элемента списка содержит: смещение центра заземлителя от проекции базовой точки в системе отсчета Натура; тип линии отрезков (окружности берут из него толщину, штриховыми не бывают); число стержней в заземлителе (от 1 до 32); угол направления отрезков заземлителя против часовой от положительного направления оси X чертежа; расстояние между стержнями (от 3000 до 25000 мм Натуры); диаметр стержней в мм Натуры.

В приведенных списках возникают связи принадлежности у объектов, отвечающих за оформление чертежа МЗС (тексты и размеры – к молниеприемникам и сечениям зон защиты, сечения зон защиты – к проекциям). Эти связи позволяют автоматизировать различные операции по модификации МЗС. Например, при удалении МП удаляются и связанные с ним тексты, радиусы и выносные линии размеров (сами размеры автоматически перегенерируются или удалятся). При переносе МП автоматически перечерчиваются учитывающие его сечения зон защиты, размеры радиусов зон защиты и др. При изменении высоты горизонтальных сечений зон защиты на виде сверху тексты этих сечений и радиусы зон защиты молниеприемников автоматически перегенерируются.

*Установки МЗС*

Установки МЗС делятся на установки, действующие по умолчанию, и общие установки. Общие установки при изменении приводят к соответствующим изменениям во всех элементах. Установки, действующие по умолчанию, применяются только при добавлении новых объектов в МЗС.

Общие установки:

- положение базовой точки внутренней системы координат (вида сверху) в чертеже, система отсчета Бумага чертежа;
- тип зоны защиты (А или Б), определяющий уровень жесткости требований к зоне защиты;
- установки вида сверху: смещение точки середины полки буквенного обозначения относительно базовой точки в системе отсчета Бумага; обозначение вида сверху; масштаб Бумага/Натура вида сверху; смещения от полки до текстов над и под ней в системе отсчета Бумага; признак добавления масштаба вида сверху к его обозначению отдельной строкой под полкой, а не в одной строке с ним; установка шрифта обозначения вида сверху.
- установки таблицы расчета молниезащиты: вектор смещения верхнего левого угла таблицы от базовой точки в системе отсчета Бумага; единица измерения выводимых в таблице числовых значений (миллиметр, сантиметр или метр); установки точности представления числовых значений в таблице (число знаков после десятичной точки, 0-7); высота строк в области данных, в мм Бумаги (3-50 мм); типы линий шапки, краев таблицы и линий-разделителей данных таблицы; установка шрифта для всех текстов таблицы в области данных (кроме текстов шапки); способ упорядочивания строк (не сортировать – без упорядочивания; сортировать – сквозное упорядочивание строк в алфавитном порядке по позиционным обозначениям МП; сортировать по группам – упорядочивание строк в алфавитном порядке по позиционным обозначениям МП, причем сначала идут все одиночные, а затем все двойные молниеприемники); признак объединения строк для одиночных молниеприемников при совпадении у них всех числовых значений;
- молниеприемников: сторона квадрата условного графического обозначения (УГО) МП на плане; диаметр жирной точки УГО МП на плане; диаметр жирной точки – сечения троса на вертикальном сечении; ширина основания треугольника УГО МП на вертикальном сечении (все в мм Бумаги);
- вертикальных сечений: установка шрифта букв "А А" буквенного обозначения на виде сверху; установка шрифта для "А – А повернуто М 1: 100" на самом вертикальном сечении; длина хвоста стрелки; длина самой стрелки; цвета обозначения сечения на виде сверху и на своем виде (все в мм Бумаги);
- текстов молниеприемников: установка шрифта и цвет;
- текстов (отметок высот) сечений зон защиты: установка шрифта; установки точности представления значения отметки высоты (число знаков после десятичной точки, 0-7) и цвет;
- общие для всех размеров: установка шрифта; длина перехода выносных за размерную (0-7.5мм); размер стрелки или засечки (0-12.6мм);
- каждого вида размера: число знаков после десятичной точки и цвет;

- заземлителей: цвет;
- для штрихования сетки для сеточного варианта МП во время работы: установка штрихования – базовая, не линейчатая.

Установки, действующие по умолчанию:
- молниеприемников: вариант конструкции (стержневой, сеточный, тросовый и двойной тросовый); отдельно стоящий МП, или устанавливается на объекте (для стержневого МП); цвет и тип линии;
- обозначений вертикальных сечений: длина штрихов на краях (0-25.5 мм); смещение стрелки от края штриха (0-25.5 мм); смещения от полки до текстов над и под ней (0-25.5 мм), все в мм Бумаги; признак добавления масштаба сечения к его буквенному обозначению отдельной строкой под полкой, а не в одной строке с ним;
- сечений зон защиты: цвет и тип линии;
- общие для всех текстов: признак соединения сноски с полкой текста (сноска на конец полки или на начало);
- текстов (отметок высот) сечений зон защиты: признак расположения текста в двух строках или в одной;
- общие для всех размеров: смещение текстов всех размеров от размерной линии в направлении допустимого расположения текстов в мм Бумаги; признак автоматического позиционирования текста размера посередине размерной линии, а не задается вручную пользователем; признак наличия сноски и указания на размерную (нет, сноска указывает на середину, начало или конец размерной линии); признак проведения сноски к началу или к концу полки текста; сведения о засечках или стрелках (стрелка внутрь, стрелка наружу или засечка);
- размеров – радиусов зон защиты одиночных стержневых МП на виде сверху: признак включения в текст значения высоты сечения зоны защиты (включать или не включать);
- заземлителей: тип линии отрезков (окружности берут из него толщину, штриховыми не бывают); число стержней в заземлителе (от 1 до 32); угол направления отрезков заземлителя против часовой стрелки от положительного направления оси X чертежа; расстояние между стержнями (от 3000 до 5000 мм Натуры); диаметр стержней в мм Натуры.

Описанное параметрическое представление МЗС может сохраняться на диске как комплект параметров, без геометрической части. Однако при выборе комплекта параметров на диске для работы с ним пользователь просматривает изображение МЗС, которое генерируется при перемещении по меню в режиме on-line.

*Операции по подготовке чертежа проекта МЗС*

Над параметрическим представлением МЗС проектировщик последовательно выполняет различные операции, выбирая их из основного меню.

"С диска" – чтение из дискового файла ранее сохраненного в нем параметрического представления МЗС.

"Из чертежа" – взятие параметрического представления из проекта МЗС в чертеже.

"Добавление" – группа операций по добавлению в модуль МЗС: молниеприемников, сечений чертежа, сечений зон защиты, текстов молниеприемников, текстов сечений зон защиты на виде сверху, размеров расстояний между осями стержневых молниеприемников, размеров радиусов зон защиты одиночных стержневых молниеприемников на виде сверху, размеров радиусов зон защиты одиночных молниеприемников на вертикальных сечениях, размеров минимальных ширин зон защиты двойных молниеприемников на виде сверху, заземлителей на виде сверху, молниеприемников, вносимых в таблицу расчета молниезащиты. Добавление объектов выбранного типа повторяется циклически до отказа от очередного добавления.

"Удаление" – запускается режим повторяющегося удаления объектов МЗС: молниеприемников, сечений чертежа, сечений зон защиты, текстов молниеприемников, текстов сечений зон защиты на виде сверху, размеров расстояний между осями стержневых молниеприемников, размеров радиусов зон защиты одиночных стержневых молниеприемников на виде сверху, размеров радиусов зон защиты одиночных молниеприемников на вертикальных сечениях, размеров минимальных ширин зон защиты двойных молниеприемников на виде сверху, заземлителей на виде сверху, молниеприемников, вносимых в таблицу расчета молниезащиты, выбираемых курсором по одному.

При удалении молниеприемника на вертикальном сечении он удаляется лишь из него (это равнозначно удалению молниеприемника из сечения зоны защиты на вертикальном сечении чертежа), на виде сверху он не удаляется.

"Перенос" – запускается циклический режим переноса объектов: молниеприемников, сечений чертежа, текстов молниеприемников, текстов сечений зон защиты на виде сверху, размерной линии размеров расстояний между осями стержневых молниеприемников, размерной линии размеров радиусов зон защиты одиночных стержневых молниеприемников на виде сверху, размерной линии размеров радиусов зон защиты одиночных молниеприемников на вертикальных сечениях, заземлителей на виде сверху, выбираемых курсором по одному.

Перенос молниеприемника на вертикальном сечении чертежа возможен только по вертикали, это равнозначно изменению высоты молниеприемника.

"Редактирование" – группа операций по редактированию объектов МЗС:
- "Добавление МП в сечение зоны защиты";
- "Удаление МП из сечения зоны защиты";
- "Добавление вершины сеточного МП";
- "Перенос вершины (опоры) МП";
- "Удаление вершины сеточного МП";
- "Перенос обозначения сечения чертежа";
- "Изменение секущего отрезка";
- "Совмещение видов";
- "Перенос точки указания сноски";
- "Перенос текста";
- "Правка данных в таблице".

"Копирование" – запускается циклический режим молниеприемников и заземлителей, выбираемых курсором по одному. Цикл работы с одним объектом выглядит так:
- в чертеже выбирается один из перечисленных выше объектов;
- создается копия объекта, изображение которой начинает перемещаться по полю чертежа вместе с курсором;
- в нужном месте чертежа нажимается клавиша Enter (левая кнопка мыши).

Цикл повторяется до нажатия Esc (правой кнопки мыши) в ситуации выбора очередного объекта или указания положения копии объекта.

"Свойства" – запускается циклический режим изменения свойств объектов МЗС: молниеприемников, сечений чертежа, сечений зон защиты, текстов молниеприемников, текстов сечений зон защиты на виде сверху, размеров расстояний между осями стержневых молниеприемников, размеров радиусов зон защиты одиночных стержневых молниеприемников на виде сверху, размеров радиусов зон защиты одиночных молниеприемников на вертикальных сечениях, размеров минимальных ширин зон защиты двойных молниеприемников на виде сверху, заземлителей на виде сверху, выбираемых курсором по одному.

"Перенос МЗС" – перенос всего проекта МЗС в другое место на поле чертежа.

"Установки" – изменение установок, действующих при подготовке МЗС:
- "Тип зоны, установки плана";
- "Молниеприемники";
- "Сечения чертежа";
- "Сечения зон защиты";
- "Тексты";
- "Тексты молниеприемников";
- "Тексты сечений зон защиты";
- "Размеры";
- "Размеры расстояний";
- "Радиусы на виде сверху";
- "Радиусы на верт.сечениях";
- "Размеры мин.ширин";
- "Заземлители";
- "Таблица расчета".

"Высота в точке" – подсветка высоты полной зоны защиты (от всех молниеприемников). Эта операция позволяет в любой точке, указываемой курсором на плане, отображать значение высоты полной зоны защиты.

"Сечение в точке" – операция идентична предыдущей, но только вместе с высотой полной зоны защиты отображается и ее горизонтальное сечение на этой высоте. Текст значения высоты и изображение сечения зоны защиты меняются вместе с движением курсора.

"Рельеф" – позволяет просмотреть "послойный" набор горизонтальных сечений полной зоны защиты, аналогичный изобазам в географических картах. Горизонтальные сечения строятся, начиная с нулевой отметки высоты и до максимальной из габаритных высот молниеприемников с шагом в 5% этого диапазона. Далее включается режим просмотра с подсветкой высоты полной зоны защиты (рис.1).

"На диск" – запись на диск текущего параметрического представления МЗС. Его можно будет повторно использовать в дальнейшем, копировать на другие рабочие места.

"В чертеж" – помещение проекта МЗС в чертеж.

*Таблица расчета молниезащиты*

Изображение таблицы расчета молниезащиты вычерчивается автоматически на основе текущего состояния МЗС, выбора одиночных и двойных молниеприемников и соответствующих им значений высот защищаемых уровней (рис.2).

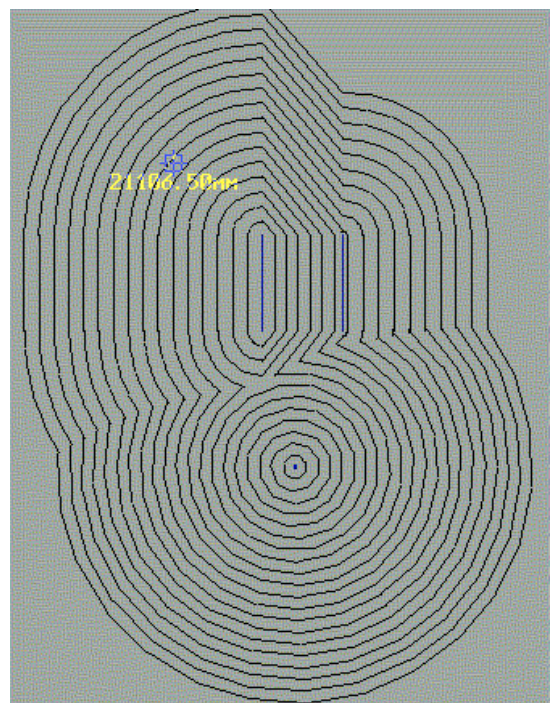

Рис.1. Рельеф полной зоны защиты

| №№ Молниеприемников | Высота молниеприемника | Активная высота молниеприемника | Высота защищаемого уровня | Радиус зоны защиты одиночного молниеприемника | Расстояние между молниеприемниками | Минимальная высота зоны защиты двух молниеприемников | Минимальная ширина защиты двух молниеприемников | Тип молниеприемника |
|---|---|---|---|---|---|---|---|---|
| | h | $h_0$ | $h_x$ | $r_x$ | L | $h_c$ | $r_{cx}$ | |
| МА-1 | 1950 | 1846 | 1750 | 1289 | | | | СМ-1 |
| МА-2, МА-3 | 1550 | 1346 | 1150 | 1089 | | | | СМ-3 |
| МА-4 | 1987 | 1888 | 1600 | 1389 | 5600 | 1750 | 2454 | СМ-2 |
| МА-5 | 1689 | 1430 | | 577 | | | | СМ-4 |

Рис.2. Таблица расчета молниезащиты

*Выводы*

Изложенные в настоящей работе состав и особенности реализации операций подготовки проекта МЗС вместе с их структурированным параметрическим представлением являются системной моделью чертежей молниезащиты зданий и сооружений, используемой при проектировании реконструкции предприятий согласно системе проектной документации для строительства. Модель разработана по модульной технологии [1] и прошла практическую апробацию в условиях проектно-конструкторского отдела крупного предприятия химической промышленности..

## Литература

[1] Мигунов В.В. Модульная технология разработки проблемно-ориентированных расширений САПР реконструкции предприятия / Материалы Второй международной электронной научно-технической конференции "Технологическая системотехника" (ТСТ'2003), г.Тула, 01.09.2003-30.10.2003 [Электронный ресурс] / Тульский государственный университет. – Режим доступа: http://www.tsu.tula.ru/aim/, свободный. – Загл. с экрана. – Яз. рус., англ.
[2] РД 34.21.122-87 "Инструкция по устройству молниезащиты зданий и сооружений". – М.: Министерство энергетики и электрификации СССР, 1988.